\documentclass[%
 reprint,
 amsmath,amssymb,
 aps,
 longbibliography,
 floatfix,
]{revtex4-1}

\usepackage{graphicx}
\usepackage{dcolumn}
\usepackage{bm}
\usepackage{hyperref}

\begin{document}


\title{Visualization of the effect of structural supermodulation on electronic structure in IrTe$_{2}$ by scanning tunneling spectroscopy}

\author{T. Machida$^{1,2}$, Y. Fujisawa$^{1}$, K. Igarashi$^1$, A. Kaneko$^1$, S. Ooi$^2$, T. Mochiku$^{2}$, M. Tachiki$^{2}$, K. Komori$^2$, K. Hirata$^{2}$ and H. Sakata$^{1}$}

\affiliation{$^{1}$Department of Physics, Tokyo University of Science, 1-3 Kagurazaka, Shinjuku-ku, Tokyo 162-8601, Japan\\
$^{2}$Superconducting Properties Unit, National Institute for Materials Science, 1-2-1 Sengen, Tsukuba, Ibaraki 305-0047, Japan}
\date{\today}

\begin{abstract}
We report on the scanning tunneling spectroscopy experiments on single crystals of IrTe$_{2}$.
A structural supermodulation and a local density-of-states (LDOS) modulation with a wave vector of $q$ = 1/5$\times$$2\pi /a_{0}$ ($a_{0}$ is the lattice constant in the $ab$-plane) have been observed at 4.2K where the sample is in the monoclinic phase.
As synchronized with the supermodulation, the LDOS spatially modulates within two energy ranges (below -200 meV and around -100 meV).
We further investigated the effect of the local perturbations including the antiphase boundaries and the twin boundaries on the LDOS.
These perturbations also modify the LDOS below -200 meV and around -100 meV, even though the lattice distortions induced by these perturbations appear to be different from those by the supermodulation.
Our results indicating several microscopic structural effects on the LDOS seem to offer crucial keys for the establishment of the microscopic model describing the parent state.
\end{abstract}

\maketitle
\begin{figure}[htbp]
\begin{center}
\includegraphics[width=7cm]{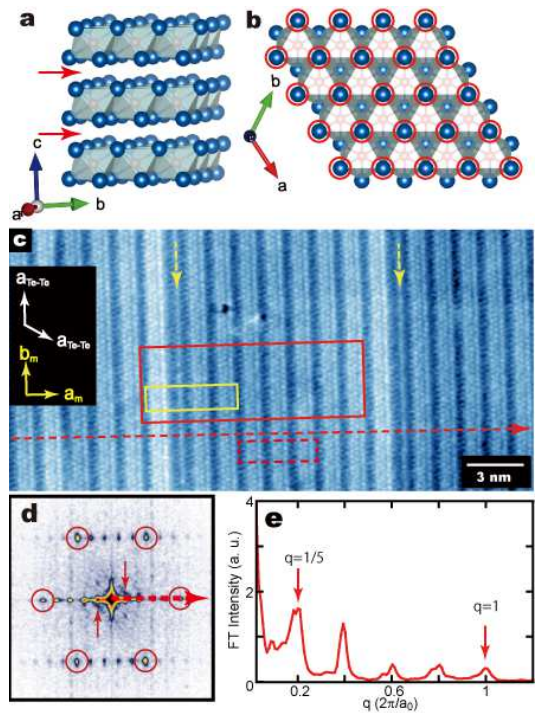}
\end{center}
\caption{(Color online)(a) and (b) Schematic illustrations of the crystal structure of IrTe$_{2}$ viewed from the direction perpendicular and parallel to the $c$-axis, respectively.
(c) A typical STM image on the surface exposed by the cleavage at 4.2 K on a 35$\times$17.5 nm$^{2}$ field of view at the bias voltage $V_{\mathrm{set}}$ = 200 mV and the current $I_{\mathrm{set}}$ = 300 pA.
In this image, white and yellow arrows represent the crystallographic orientations of the high-temperature trigonal and the low-temperature monoclinic phases, respectively.
Yellow dashed arrows indicate the position of the anti-phase boundary.
(d) Fourier transform image of (c).
Red circles correspond to the Fourier peaks attributed to the triangular lattice composed of the topmost Te atoms.
Red arrows indicate the spots corresponding to the structural supermodulation.
(e) Linecut of the FT intensity along red dashed arrow in (d).
}
\label{Fig_1}
\end{figure}

\begin{figure*}[htbp]
\begin{center}
\includegraphics[width=14cm]{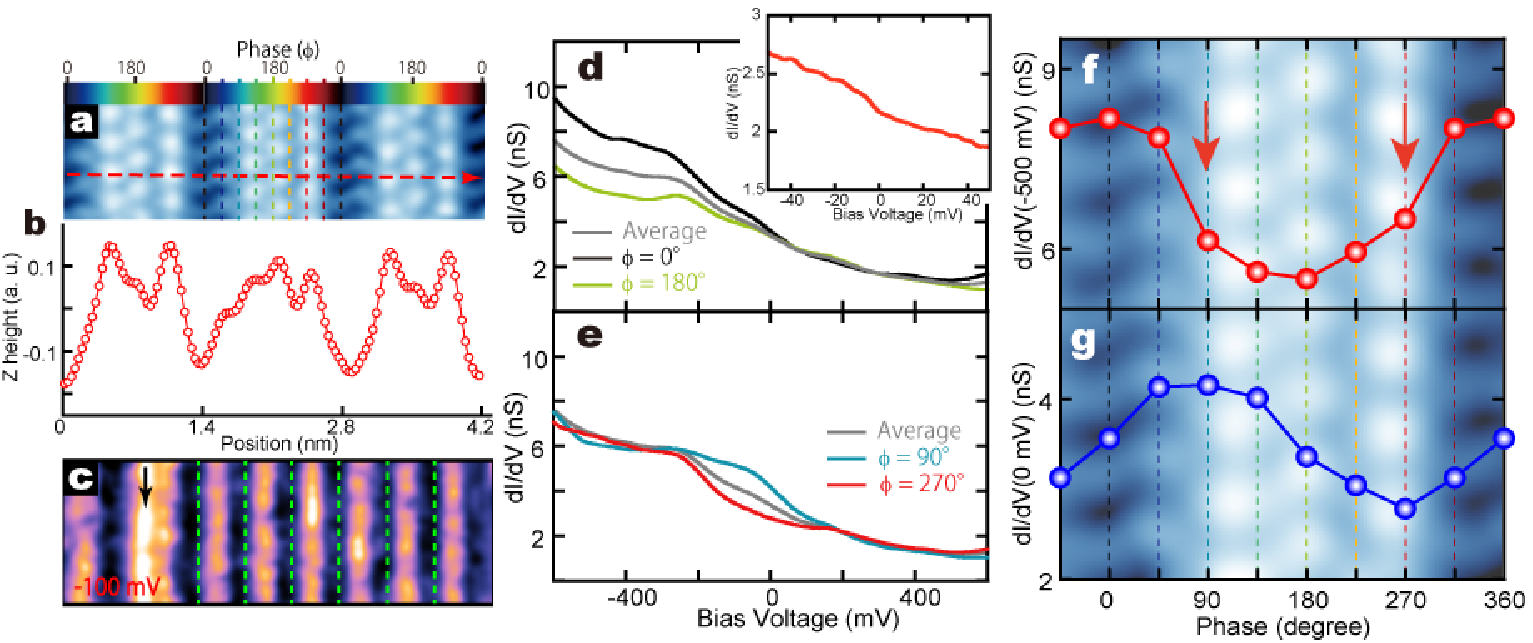}
\end{center}
\caption{(Color online)
(a) A magnified STM image taken on the region marked by red dashed box in Fig. 1(c) (4.3 $\times$ 1.2 nm$^2$). Color scale and colored dashed lines represent the phase ($\phi$) of the supermodulation, which runs along the transverse direction in this image.
Here, we define the phase ($\phi$) as $\phi$ = 0 at the trough of the supermodulation.
(b) Line profile of the topography along red dashed arrow in (a).
(c) Conductance map at -100 mV taken on the region marked by red box in Fig. 1(c). Green dashed lines correspond to the trough of the supermodulation.
(d) Black and green lines are the spectra taken at the trough ($\phi$ = 0$^{\circ}$) and the crest ($\phi$ = 180$^{\circ}$) of the modulation, respectively. Inset indicates the representative tunneling spectrum taken at high energy resolution ($\sim$2 meV).
(e) In this graph, the spectra taken at the two midpoints between the crest and trough of the modulation are represented by light blue ($\phi$ = 90$^{\circ}$) and red ($\phi$ = 270$^{\circ}$) lines, respectively.
In (d) and (e), the spatially averaged tunneling spectra are shown by gray lines.
(f) and (g), phase (position) dependence of the conductance at -500 mV (f) and at 0 mV (g).
}
\label{Fig_2}
\end{figure*}

\section{Introduction}
Materials containing transition metals are often rich in interesting phenomena induced by the peculiar nature of the $d$ or $f$ states of the transition metals.
Intriguing examples are several types of charge density wave (CDW) state in transition metal chalcogenides\cite{Hess,Guillamon,Brun_1,Sipos,AFang,JJKim}, electronic liquid crystal, electronic stripe and high-temperature superconductivity in Fe- and Cu-based superconductors (SCs) \cite{Lawler,Mesaros,TMChuang,Kasahara}.
Particularly, mysterious phenomena in Fe-based superconductors, including the high-temperature superconductivity and the electronic nematicity, are intimately tied to the Fe 3$d$ orbital degree of freedom and have attracted much attention in condensed matter physics.
Recently, in Ir$_{1-x}$X$_{x}$Te$_{2}$ (X = Pt or Pd), several similar phenomena to those observed in Fe-based SCs have been discovered; (i) the occurrence of the structural phase transition and the suggestion of the orbital order in undoped parent compound of IrTe$_{2}$, and (ii) the emergence of the superconductivity with the chemical substitution and intercalation which suppresse the structural transition \cite{SPyon,JJYang,Kamitani}.
Due to these similarities, the understanding of the relation between the structural transition, orbital order and superconductivity in Ir-based SCs would give crucial hints to uncover the long-standing issues regarding the mechanism of the unconventional superconductivity.
As a first step, it is important to unveil what is occurring in the parent compound of IrTe$_2$.

Recent electron diffraction study in undoped IrTe$_2$ indicated that the diffraction peak with $q$=(1/5, 0, -1/5) appears below structural transition temperature.
This suggests the existence of a structural supermodulation with the wave vector $q$=(1/5, 0, -1/5).
In addition to the supermodulation, Yang \textit{et al.} predicted that a charge modulation emerges with the same period and along the same direction as those of the supermodulation below the structural transition temperature\cite{JJYang}.
They have concluded that the structural transition stems from the orbitally induced Peierls mechanism\cite{Khomskii}.
However, an optical spectroscopy and ARPES studies exhibited no signature of the energy gap near $E_{\mathrm{F}}$ which is hallmark of a CDW associated with the Peierls transition\cite{AFFang,Ootsuki_2}.
Additionally, the recent band calculations have suggested that the structural transition is driven by the crystal field effect mainly from the Te 5$p$ orbital splitting\cite{AFFang,Kamitani}.
More recently, it has been also suggested that the depolymerization of the polymeric Te-Te bonds is responsible for the structural transition\cite{YS_Oh}.
Thus, it has been still controversial what is the origin of the structural transition and how the parent state is related to the superconductivity.
These issues will be solved by the establishment of the microscopic model in the parent state, which is difficult due to the complexity induced by the existence of the supermodulation.
Therefore, it is quite crucial to visualize the effect of the supermodulation on the electronic states.

In this study, we present the scanning tunneling microscopy (STM) studies on single crystals of IrTe$_{2}$.
The results indicate the existence of a supermodulation and an LDOS modulation with a wave vector of $q$ = 1/5$\times$$2\pi /a_{0}$ ($a_{0}$ is the lattice constant in the $ab$-plane).
There are two energy ranges within which the LDOS spatially varies as synchronized with the supermodulation.
Besides the supermodulation, the antiphase boundaries and the twin boundaries also modify the LDOS below -200 meV and around -100 meV, even though the lattice distortions induced by these perturbations appear to be different from those by the supermodulation.
These microscopic structural effects on the LDOS may be crucial for the establishment of the microscopic model describing the parent state of this material.

\section{Experimental details}
Single crystals of IrTe$_{2}$ used in this study were grown by a self-flux method\cite{JJYang}.
The structural transition temperature was determined to be approximately 280 K (heating process) by the electric transport measurements.
We used a laboratory-built cryogenic STM for our scanning tunneling spectroscopy (STS) measurements.
All of the measurements were carried out at 4.2 K.
Samples were cleaved \textit{in}-\textit{situ} at 4.2 K.
The cleavage exposes the surface composed of the Te atoms due to the weak van der Waals coupling between two Te layers as shown in Fig. 1(a) and (b).
An electrochemically polished Au wire was used as an STM tip.
Before proceeding to measurements on a sample, we scanned on an Au thin film (200 nm thickness) deposited on a cleaved surface of mica to verify the quality of the STM tip.
The STM topographic images were obtained in constant-current mode.
The $dI/dV$ conductance spectra were obtained by numerical differentiation of the $I$-$V$ characteristics measured at each location.

\section{Results and discussion}
\subsection{Supermodulation with $q = 1/5(2\pi/a_{0})$}
A typical topographic image contains an approximately triangular lattice with a period of $\sim$ 3.9 \AA\ corresponding to the distance between topmost Te atoms, as shown in Fig. 1(c) and Fig. 2(a).
The structure modulates periodically along one of the Ir-Te bond directions with a wave vector of $q$ = 1/5$\times$$2\pi /a_{0}$, which is confirmed by a Fourier transform image as shown in Fig. 1(d) and (e).
The observed supermodulation corresponds to the projection on the $c$-plane [(001) plane] of the previously observed supermodulation with the wave vector $q$=(1/5, 0, -1/5) \cite{JJYang}.
In Fig. 2(b), we plotted a line profile of the topography parallel to the monoclinic $a$-axis, indicating that the shape of the observed supermodulation is non-sinusoidal form as suggested by the previous TEM measurements\cite{YS_Oh}.
In this work, we choose to label the phase of the supermodulation as $\phi$ = 0$^{\circ}$ (180$^{\circ}$) at the trough (crest) of the supermodulation.

\subsection{Local density of states modulation}
We first focus on the energy dependence of the local density-of-states (LDOS).
Figure 2(d) and (e) show the representative tunneling spectra exhibiting a particle-hole asymmetry characterized by the higher conductance in negative energy than that in positive.
This particle-hole asymmetry is qualitatively consistent with the band calculations\cite{JJYang,Ootsuki_1,AFFang}.
It is noted that the energy resolution of the spectra in Fig. 2(d) and (e) is about 20 meV.
We also observed the tunneling spectra with the higher energy resolution that is a few meV as shown in the inset of Fig. 2(d) and did not find an decisive energy gap near Fermi energy except an irreproducible gap structure about 10 meV occasionally observed.

To explore the effect of the supermodulation on the electronic states, we compare the tunneling spectra taken at the locations having different phases of the supermodulation.
A distinct difference between the spectra at the crest and trough appears below -200 mV whereas no remarkable difference is observed in the positive energy: the conductance below -200 mV is suppressed (enhanced) at the crest $\phi$ = 180$^{\circ}$ (trough $\phi$ = 0$^{\circ}$) of the supermodulation as shown in Fig. 2(d).
To reveal how the conductance below -200 mV changes spatially, we plot the conductance at -500 mV as a function of the phase as in Fig. 2(f).
There are two features in the spatial variation of the conductance below -200 mV.
First one is that the phase of the conductance modulation is completely opposite to that of the supermodulation.
Second is abrupt changes around $\phi$ = 90$^{\circ}$ and 270$^{\circ}$ where the topography suddenly changes, as indicated by the arrows in Fig. 2(f).
Figure 2(e) shows the spectra taken at the two midpoints between the crest and trough ($\phi$ = 90$^{\circ}$ and 270$^{\circ}$).
The conductances in both spectra are unchanged below -200 mV and above +100 mV.
The significant difference lies around -100 mV.
The conductance modulation around -100 mV has a relative phase difference with respect to the conductance modulation below -200 mV as shown in Fig. 2(g): the local maximum and minimum appear at around 90$^{\circ}$ and 270$^{\circ}$.
Thus, it is conceivable that there are the two energy scales (below -200 meV and around -100 meV) in which the LDOS spatially modulates with structural change due to the supermodulation.
The spatial variations of the LDOS in the two energy scales are quite different with each other.

\begin{figure*}[htbp]
\begin{center}
\includegraphics[width=12cm]{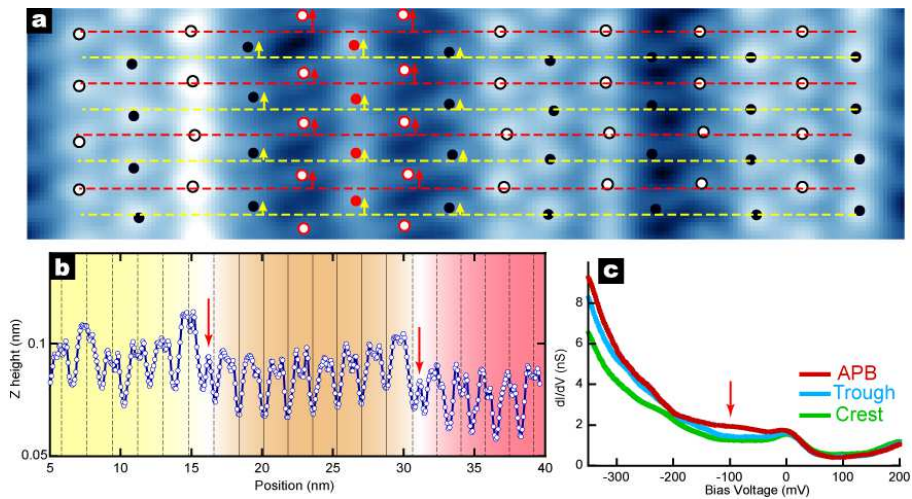}
\end{center}
\caption{(Color online)
(a) A magnified STM image taken on the field of view indicated by yellow box in Fig. 1(c). Red solid and opened circles correspond to the Te atoms on the APB and the nearest neighbor Te atoms, respectively.
Black solid and opened circles indicate other Te atoms not marked by red solid and opened circles.
Red and yellow dashed lines display the averaged vertical positions of the black opened and solid circles.
Arrows show the shift of the Te atoms around the APB from the averaged position.
(b) Line profile of the topography along the red dashed arrow in Fig. 1(c) including the two APB indicated by two red arrows.
(c) Tunneling spectra taken at the APB (red), the trough (light blue) and the crest (light green) of the supermodulation. The set up parameters in taking these spectra are $V_{\mathrm{set}}$ = 500 mV and $I_{\mathrm{set}}$ = 500 pA).
}
\label{Fig_3}
\end{figure*}

Here we discuss the origin of the spatial variations of the LDOS in the two energy scales. In the several previous works, it has been suggested that the distortions of Ir-Ir, Ir-Te, intralayer Te-Te, and interlayer Te-Te bonding affect the electronic structure via the orbital degree of freedom of Ir 5d and Te 5p orbitals \cite{JJYang,Ootsuki_1,Ootsuki_2,AFFang,Kamitani,YS_Oh,HB_Cao,GLPascut}. Particularly, recent first principle calculation based on the new crystal symmetry determined by thorough X-ray measurements has revealed that the LDOS (site dependence of DOS) drastically varies across the supermodulation. According to the calculation, the dimerization of the two out of the five Ir atoms creates the anti-bonding and bonding states of Ir $d_{\mathrm{xy}}$ orbitals around -3 eV and + 1eV at the position of the Ir-Ir dimer.  Additionally, the energy asymmetry of the DOS within the energy range from -500 to +500 meV is suppressed at the position of the Ir-Ir dimer. This site dependence of the energy asymmetry of the DOS is qualitatively consistent with our results. If this calculation is valid, the position of the dimer corresponds to the brightest area (around $\phi$ = 180 $^\circ$) in our STM image of Fig. 2(a) or (f), because the asymmetry of the tunneling spectrum is suppressed at the crest of the supermodulation.
Thus, the Ir-Ir dimerization is one of possible origins of the observed dI/dV modulation below -200 mV.

Even if we use this calculation results, we cannot clearly explain the observed phase difference between the
dI/dV modulation below -200 and around -100 mV.
One possible origin of the observed phase difference is the phase difference between the supermodulation in an IrTe$_{2}$ layer and that in its neighboring layer.
This phase difference produces a periodic modulation of the interlayer Te-Te distance ($d_{\mathrm{Te}}$) with a different phase as that of the supermodulation \cite{HB_Cao,GLPascut}.
Such the periodic modulation of $d_{\mathrm{Te}}$ seems to be able to modify periodically the electronic structure, following the previous suggestion that the $d_{\mathrm{Te}}$ drastically affects the electronic band dispersion around $E_{\mathrm{F}}$ \cite{Kamitani}.
If this periodic modulation of $d_{\mathrm{Te}}$ really creates the spatial variation of the LDOS, there would be a finite phase difference between the LDOS modulations due to the periodic modulation of $d_{\mathrm{Te}}$ and due to the Ir-Ir dimerization, because the there is a finite phase difference between the position of the Ir-Ir dimer and the modulation of $d_{\mathrm{Te}}$.
Thus, it is presumable that there are two components governing the LDOS: (i) the Ir-Ir dimerization, which affects mainly the energy asymmetry of the LDOS from -500 to +500 meV, (ii) the interlayer Te-Te distance, which governs the LDOS around -100 meV.

\subsection{Effects of anti-phase boundary and twin boundary on LDOS}
Besides the effect of the supermodulation on the LDOS, we also investigated the influences of the local perturbations including antiphase boundaries (APBs) and twin boundaries (TB) which are previously reported \cite{YS_Oh}.
As shown in Fig. 1(c), we sometimes found several APBs at which the phase of the supermodulation varies.
In the case of Fig. 1(c) containing two APBs, the phase shifts across these boundaries is approximately 6$\pi$/5, as shown in Fig. 3(a) and (b).
When such APBs appear periodically, the wave vector of the supermodulation should be incommensurate \cite{Soumyanarayanan}.
However, the wave vector measured by the previous TEM experiments\cite{JJYang,YS_Oh} is commensurate at least their experimental accuracy.
Therefore, it is presumable that the observed APBs localize and inhomogeneously distribute in the crystal.
As shown in Fig. 3(a), we can find slight shifts of the topmost Te atoms along the $b_{\mathrm{m}}$-axis [vertical direction in Fig. 3(a)].
In the tunneling spectra around the APB inducing such the lattice distortion, the conductance around -100 mV and below -200 mV are slightly enhanced, as shown in Fig. 3(c).

In addition to the APBs, we rarely found twin boundaries at which the crystal orientation changes by 60$^{\circ}$.
There are several types of the twin boundaries in our crystals as shown in Fig. 4(a) and (b) where the representative two boundaries are displayed.
The height is enhanced (suppressed) at the boundaries in Fig. 4(a) [Fig. 4(b)].
At the boundary where some lattice distortions reside, the conductance suppression around -100 mV and enhancement below -200 mV can be seen as shown in Fig. 4(c) and (d).
Thus, although the lattice distortions induced by these perturbations seem to be different from those by the supermodulation, LDOS in the same energy ranges respond to the distortions.
Therefore, there seems to be a similar mechanism linking between these distortions and their effects on the LDOS.

\begin{figure*}[htbp]
\begin{center}
\includegraphics[width=13cm]{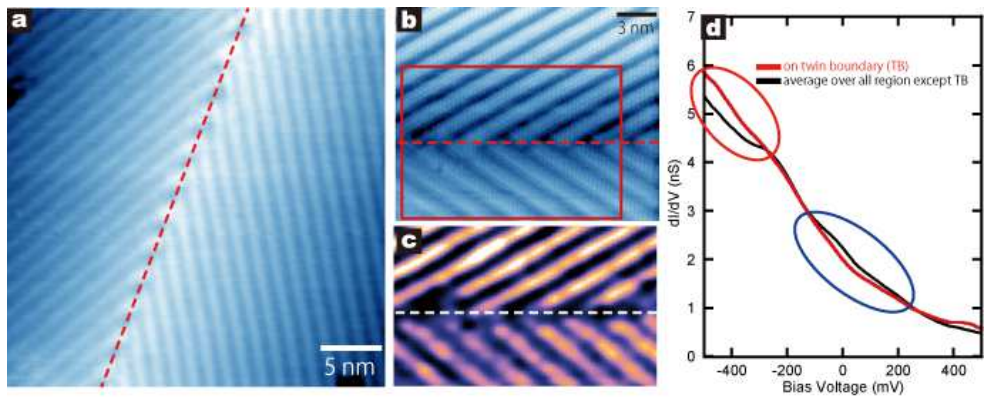}
\end{center}
\caption{(Color online)
(a) and (b), Typical STM images of two types of TBs. (a) is taken on a 31 $\times$ 31 nm$^2$ field of view at $V_{\mathrm{set}}$ = 200 mV and $I_{\mathrm{set}}$ = 600 pA. (b) is taken on a 20 $\times$ 16.3 nm$^2$ field of view at $V_{\mathrm{set}}$ = 500 mV and $I_{\mathrm{set}}$ = 1 nA.
Red dashed lines correspond to the boundaries in these images.
(c) A conductance map at -100 meV on the region indicated by red box in (b).
(d) Tunneling spectra taken at the boundary (red) and averaged over all region except the TB ($V_{\mathrm{set}}$ = 500 mV, $I_{\mathrm{set}}$ = 1 nA.) .
}
\label{Fig_4}
\end{figure*}

\section{Summary}
In summary, we have performed the scanning tunneling spectroscopy experiments on IrTe$_{2}$ at 4.2K where the samples are in the low temperature monoclinic state.
The results indicate a structural supermodulation and an LDOS modulation with a wave vector of $q$ = 1/5$\times$$2\pi /a_{0}$.
There are two energy ranges where the LDOS sensitively respond to the supermodulation: (i) below -200 meV and (ii) around -100 meV.
We further investigated the effect of local perturbations i.e. the APB and TB on the LDOS.
Even though the actual lattice distortions induced by these perturbations appear to be different from those by the supermodulation, the LDOS are also modified by these perturbations below -200 meV and around -100 meV as seen in the supermodulation.
These microscopic structural effects on the LDOS seem to be crucial for the establishment of the microscopic model describing the parent state of this material which is necessary to understand the relation between the parent and superconducting states and to uncover the origin of the structural transition.

\end{document}